# Efficient Method for Updating IDF Curves to Future Climate Projections

Jonathan B. Butcher, M.ASCE, Ph.D., P.H.[1]; Tan Zi, Ph.D., P.E.[2]

1. Hydrologist, Tetra Tech, Inc., Research Triangle Park, NC 27709 (corresponding author).

   E-mail: jon.butcher@tetratech.com.

2. Environmental Engineer, Tetra Tech, Inc., Fairfax, VA 22030

**Abstract**

If climate stationarity is dead, how should engineering design standards be modified to account for potential changes in extreme precipitation? Many standards rely on precipitation intensity-duration-frequency (IDF) curves provided in NOAA's Atlas 14. General Circulation Models (GCMs) predict increases in average temperature throughout the US, but are less clear on changes in precipitation. In many areas GCMs predict relatively small changes in total precipitation volume, but also suggest increased magnitude of extreme events as warmer air can hold more water. Unfortunately, GCMs have limited skill in predicting individual storm events. This research developed an efficient approach to solve this problem that replicates Atlas 14 methods, which fit a generalized extreme value distribution to the annual maximum series. GCM biases are addressed through equidistant quantile mapping, mapping from a GCM's historical simulation to future predictions, then applying this mapping from the fit to data used in Atlas 14. The approach is efficient because it requires only the annual maxima, allowing rapid examination of results across the range of GCM projections.

**Introduction**

Engineering design for stormwater management is largely based on empirical evidence obtained from past data with the assumption that the frequency of extreme events that is likely to be seen in the future can be inferred from the historical record. This implies that climate is stationary. However, predicted changes in future climate imply the end of the assumption of stationarity that has provided the foundation of water management for decades, as was announced by Milly et al. (2008). Commenting on the "death of stationarity", Galloway (2011) noted "there is also a great need to provide those in the field the information they require now to plan, design, and operate today's projects."

Intensity-Duration-Frequency (IDF) curves graphically summarize the relationship between precipitation intensity and the duration of precipitation events for a given frequency or recurrence interval. IDF curves provide important information for engineering design and planning purposes. From one perspective, updating IDF curves for future climate is simple – conditional on estimates of the distribution of future precipitation events. Unfortunately, the skill of GCMs in predicting individual precipitation events is limited, especially convective storm events that provide the most intense storms, yet occur at spatial scales smaller than the resolution of GCMs. Sun et al. (2006) summarized reanalysis studies of GCMs as follows: "For light precipitation, most of the models greatly overestimate the frequency but reproduce the observed patterns of intensity relatively well. For heavy precipitation, most of the models roughly reproduce the observed frequency, but underestimate the intensity." Some of the biases inherent in GCMs are resolved by downscaling results to a finer, local scale, often with a bias correction step. However, Maraun et al. (2010) conclude that serious deficiencies remain in the ability of downscaling methods to generate local precipitation series with the correct temporal variability.

In the U.S., official estimates of precipitation frequency for specific geographic locations are provided as IDF curves and tables in NOAA's Atlas 14 (Perica et al., 2013). The specific objective of this work is to provide a method to update Atlas 14 IDF curves to reflect potential future changes in local climate. To satisfy this objective it is important to understand the way in which the Atlas 14 estimates were created. Specifically, frequency estimates in the Atlas are based on fitting a generalized extreme value (GEV) distribution to the time series of annual maximum precipitation (AMP) amounts at a station for seventeen durations ranging from 15 minutes to 60 days. The AMP series consists of one measurement per year, and does not account for the possibility of more than one event in a year exceeding a threshold of interest. The true probability of occurrence of events of a given intensity and duration should be derived from the partial duration series, which includes all events of a specified duration and above a pre-defined volume threshold. Frequency estimates for partial duration series were developed by NOAA for Atlas 14 from the series of AMPs using Langbein's conversion formula, which transforms a partial duration series-based average recurrence interval (ARI) to an annual exceedance probability (AEP):

$$AEP = 1 - \exp\left(-\frac{1}{ARI}\right) \qquad [1]$$

Selected partial duration ARIs are first converted to AEPs using this formula, and frequency estimates were then calculated for the AEP using the GEV fit to annual maxima.

NOAA fit the GEV for each station using the method of L-moments (Hosking and Wallis, 1997), incorporating regionalization across approximately the 10 nearest stations for higher order L-moments. NOAA does not release the fitted coefficients of the GEV distribution, although the annual maximum series are provided via ftp server. It is important to note that the NOAA method is ultimately based only on annual maxima (the AMP series). This means that only the annual maximum series is needed for future climate conditions and not the complete or partial duration series. This has important implications for the mathematical approach to updating the IDF curves, as described below.

Updating the IDF curves in Atlas 14 for future climate requires understanding how the extreme value distribution fit to annual maximum precipitation series may change. Researchers have proposed a variety of statistical and parametric methods for updating IDF curves based on GCM output (e.g., Prodanovic and Simonovic, 2007; Huard et al., 2010; Arnbjerg-Nielsen, 2013). These methods have generally been characterized by a high level of complexity and computational intensity, and have worked with complete future downscaled precipitation series, even though the Atlas 14 IDF curves depend only on the AMP series.

A simple, more direct, and computationally efficient approach to updating IDF curves was recently proposed by Srivastav et al. (2014a, 2014b). Their insight was that the essence of the problem was the need to update extreme value distributions for future conditions, and that this could be done through a direct analysis of the distributions. The general concept of the approach of Srivastav et al. (2014a) is described as follows: "…quantile-mapping functions can be directly applied to establish the statistical relationship between the AMPs of a GCM and sub-daily observed data rather than using complete records. Further, the IDF is a distributional function; therefore it would be easy to derive the functional relationships between the distributions of the GCM AMPs and sub-daily observed data. One way of deriving such relationship is by using quantile-mapping functions."

Quantile mapping (QM) methods, otherwise known as cumulative distribution function (CDF) matching methods, have long been used as a method to correct for local biases in GCM output. The method first establishes a statistical relationship or transfer function between model outputs and historical observations, then applies the transfer function to future model projections (Panofsky and Brier, 1968) and has been successfully used as a downscaling method in various climate impact studies (e.g., Hayhoe et al., 2004).

Using the notation of Li et al. (2010), for a climate variable $x$, the QM method for finding the bias-adjusted future value of a climate variable can be written as:

$$\hat{x}_{m-p.adjst.} = F_{o-c}^{-1}\left(F_{m-c}(x_{m-p})\right) \qquad [2]$$

where $F$ is the CDF of either the observations (o) or model (m) for observed current climate (c) or future projected climate (p), and $F^{-1}$ is the inverse of the cumulative distribution function. The bias correction for a future period is thus done by finding the corresponding percentile values for these future projection points in the CDF of the model for current observations, then locating the observed values for the same CDF values of the observations.

A significant weakness of the QM method is that it assumes that the climate CDF does not change much over time, and that, as the mean changes, the variance and skew do not change, which is likely not true (e.g., Milly et al., 2008). To address these issues, Li et al. (2010) proposed the equidistant quantile mapping (EQM) method, which incorporates additional information from the CDF of the model projection. The method assumes that the difference between the model and observed value during the current calibration period also applies to the future period; however, the difference between the shape of the CDFs for the future and historic periods is also taken into account. This is written as:

$$\hat{x}_{m-p.adjst.} = x_{m-p} + F_{o-c}^{-1}\left(F_{m-p}(x_{m-p})\right) - F_{m-c}^{-1}\left(F_{m-p}(x_{m-p})\right), \quad [3]$$

where the form and parameters of the CDF are not yet specified. Srivastav et al. (2014a) argue for using EQM to update IDF curves; however, the specific method of Srivastav et al. (2014b) is not directly applicable to updating Atlas 14 IDF curves in the US for several reasons:

- Canada assumes that the AMP series follows a Gumbel, rather than a GEV distribution.

- Bias-corrected statistically downscaled climate model output is not widely available for Canada, therefore the Srivastav method must also incorporate a spatial downscaling step from the coarse scale of GCMs, whereas output that is already spatially downscaled to a fine resolution grid is readily available for the US.

- The method of Srivastav et al. justifies use of EQM, but largely consists of a multi-step QM procedure, without the additional EQM corrections.

To address these issues this paper re-derives an EQM method that is consistent with U.S. design guidelines and makes use of statistically downscaled climate data readily available from GCM output.

**Methods**

A combination of EQM and QM approaches are used to update IDF curves for any location conditional on output of GCMs for future climate conditions, implemented in Python code. Two distribution mapping steps are needed to update IDF curves. The process begins with GCM output that has already been subject to spatial bias correction and downscaling to a 4x4 km spatial scale and daily time step. The first calculation step consists of additional spatial downscaling from the 4x4 km grid to the specific location of the first-order weather station used by Atlas 14 along with bias correction for the AMP series (as distinct from the general bias correction of the complete precipitation series) using the EQM method.  The second step involves temporal downscaling from daily to sub-daily durations using the QM method (EQM is not needed for this step because it does not involve bias correction).

For the first step, the historical data are the historical AMP series used by Atlas 14 ($X_{max}^{STN}$). Model data include the predicted AMP series for the same historical period ($X_{max}^{GCM}$) and for the future period of interest ($X_{max}^{GCM_{FUT}}$).  A GEV distribution is fit to each of these series, using the L-moments method (Hosking and Wallis, 1997; implemented in Python in lmoments v. 0.2.3 at https://pypi.org/pypi/lmoments/), consistent with Atlas 14 methods:

$$Prob(STN) = f\left(\theta^{STN}, X_{max}^{STN}\right) \qquad [4]$$

$$Prob(GCM) = f(\theta^{GCM}, X_{max}^{GCM}) \qquad [5]$$

$$Prob(GCM_{FUT}) = f\left(\theta^{GCM_{FUT}}, X_{max}^{GCM_{FUT}}\right) \qquad [6]$$

where $f(\ )$ is the probability density function, and $\theta$ represents the vector of parameters of the fitted distribution (GEV for this case).

To apply the EQM method, quantiles of modeled future daily extreme data are matched to the distribution for historical AMPs.  For a given percentile, it is assumed that the difference between the model and observed value also applies to the future period.  There are two EQM factors.  The first is:

$$Y_{max}^{adj1} = F^{-1}((F\left(X_{max}^{GCM_{FUT}}|\theta^{GCM_{FUT}}\right))|\theta^{STN\_daily}), \quad [7]$$

where the vertical bar "|" indicates conditional dependence, i.e., $F\left(X_{max}^{GCM_{FUT}}|\theta^{GCM_{FUT}}\right)$ indicates the cumulative distribution function of the future GCM AMP series calculated at the cumulative probability corresponding to $X_{max}^{GCM_{FUT}}$ using the parameter set $\theta^{GCM_{FUT}}$ calculated for that future

series.  To account for the difference between the CDFs for the model outputs of future and current periods, a second adjustment factor is calculated:

$$Y_{max}^{adj2} = F_{m-c}^{-1}\left(F_{m-p}(x_{m-p})\right) = F^{-1}((F(X_{max}^{GCM_{FUT}}|\boldsymbol{\theta}^{GCM_{FUT}}))|\boldsymbol{\theta}^{GCM}) \qquad [8]$$

The projected AMP series is then calculated as:

$$Y_{max}^{STN_{FUT_{daily}}} = X_{max}^{GCM_{FUT}} + Y_{max}^{adj1} - Y_{max}^{adj2} \qquad [9]$$

Once this series is calculated, a GEV fit is applied to estimate the full distribution of the 24-hour duration events.

The second step in adjusting the IDF curves is temporal downscaling to convert future daily extremes into sub-daily extremes.  The QM method was used for this purpose:  First find the corresponding percentile values for these future projection points in the CDF of the model for the historical period, then locate the observed values for the same CDF values of the sub-daily observations.  For rainfall duration *i*:

$$Y_{max,i}^{STN_{FUT_{subdaily}}} = F^{-1}\left(\left(F\left(Y_{max}^{STN_{FUT_{daily}}}|\boldsymbol{\theta}^{STN_{daily}}\right)|\boldsymbol{\theta}_i^{STN_{subdaily}}\right)\right) \quad [10]$$

As noted in Atlas 14 (Perica et al., 2013), estimates for shorter durations can be noisy due to limited data availability and are improved by smoothing.  The projected future sub-daily extreme values are thus smoothed by fitting them to a linear regression relative to the daily maximum series:

$$Y_{max}^{STN_{FUT_{subdaily}}} = a_1 * Y_{max}^{STN_{FUT_{daily}}} + b_1 \qquad [11]$$

After $Y_{max}^{STN_{FUT}subdaily}$ is calculated, GEV distributions are used to fit the extremes projections. Then rainfall amounts at certain probabilities are extracted to create the rainfall series with different recurrence intervals:

$$Y_{max,i,p}^{STN_{FUT}} = F^{-1}\left(1 - p, \left(F\left(Y_{max,i}^{STN_{FUT}}\right)\right)\right), \qquad [12]$$

where $p$ is the AEP corresponding to the desired recurrence interval.

An additional normalization step is applied to ensure consistency with Atlas 14: The changing ratio of rainfall extremes between current and future can be derived by comparing the derived historical rainfall extremes at different reoccurrence intervals based on observation data $X_{max,i,p}^{STN}$ with the spatial and temporal downscaled GCM rainfall extremes data $Y_{max,i,p}^{STN_{FUT}}$:

$$\mathbf{R}_{i,p}^{j} = \frac{Y_{max,i,p}^{STN_{FUT},j}}{X_{max,i,p}^{STN}}, \qquad [13]$$

where $j$ represents the underlying GCM. The Atlas 14 IDF curves are then updated by multiplying the published results by the appropriate ratios for the different GCMs, preserving the regional representation of higher L moments incorporated in the original Atlas 14 calculations.

For durations less than one hour, Atlas 14 (Perica et al., 2013) relates 15 and 30-minute durations to the 60-minute duration using local scaling factors:

$$\mathbf{S} = \frac{X_{k}^{Atlas\_14}}{X_{60-mintue}^{Atlas\_14}} \quad , \text{k=15-minute or 30 minute. } [14]$$

Fixed scaling factors are used in Atlas 14 for deriving 10-minute and 5-minute annual maxima. The ratio of the10-minute annual maximum to the 15-minute annual maximum is assumed to be 0.82 in Atlas 14 and the ratio for the 5-minute annual maxima is 0.57. These same assumptions can be applied to future climate conditions.

**Results**

The methods described above were tested in an application for the city of Grand Rapids, MI.  In 2013, a Climate Resiliency Report for Grand Rapids was prepared by the West Michigan Environmental Action Council (WMEAC, 2013).  This report used the Model for the Assessment of Greenhouse-gas Induced Climate Change and Regional Scenario Generator (MAGICC/SCENGEN; Wigley, 2008) to forecast potential changes in annual and seasonal temperature and precipitation in Grand Rapids through 2042.  This tool produces results at a rather coarse spatial scale of 2.5°x2.5° (about 175 x 175 miles at the latitude of Grand Rapids) and at a seasonal temporal scale.  The Resiliency Report notes "MAGICC/SCENGEN has a significant limitation in that it deals in seasonal averages and is not capable of forecasting extreme weather trends."  Both the spatial and temporal scale of this product make it insufficient for evaluating changes in IDF curves for precipitation.

Since production of Grand Rapids' Resiliency Report, the IPCC (2013) has released the 5[th] Assessment Report, incorporating results from a new round of GCM runs (Coupled Model Intercomparison Project 5 or CMIP5).  CMIP5 incorporates a number of refinements to the GCMs.  It also uses a different set of greenhouse gas concentration scenarios than the emissions-based scenarios that were used in CMIP3.  These new greenhouse gas scenarios are referred to as Representative Concentration Pathways (RCPs) and are based on a future target radiative forcing (e.g., RCP 4.5 represents radiative forcing of 4.5 $W/m^2$ in year 2100) rather than inferring the radiative forcing from uncertain projections of future population growth, energy use patterns, and associated greenhouse gas emissions.

The new CMIP5 model results were used for the purpose of updating the potential range of future IDF curves.  In addition to incorporating the latest model updates, the CMIP5 results are now available in a variety of online repositories that enable rapid screening of the range of potential future outcomes predicted by the suite of GCMs. There is also a desire, however, to maintain consistency with the Resiliency Report. The analysis in the Grand Rapids Resiliency Report used the CMIP3 emissions scenario known as A1B, which was a middle-of-the-road emissions scenario incorporating "balanced emphasis on all energy sources." There is not an exact match to this scenario in CMIP5; however, the projected greenhouse gas trajectory under A1B is bounded above and below by RCP 8.5 and RCP 4.5.

RCP 8.5 includes higher greenhouse gas concentrations, and thus greater radiative forcing and higher global atmospheric temperatures than RCP 4.5; however, the difference among individual GCMs is generally greater than the difference between RCP 4.5 and RCP 8.5 projections through at least the middle of the 21st century. Further, the greatest impacts on precipitation do not necessarily line up with increases in temperature, and for various GCMs the increase in total rainfall volume is greater with RCP 4.5 than with RCP 8.5 for the 2050-2070 period. Nonetheless, the Clausius-Clapeyron relationship establishes that the potential density of fully saturated water vapor increases with increasing temperature, leading to the potential for more intense precipitation extremes (Trenberth et al., 2003; Kao and Ganguly, 2011). For illustrative purposes this paper presents results near the median extremes of the distribution of available GCMs.

### *GCM Selection*

Future climate projections are uncertain and are best used to describe a probability envelope of potential future conditions (an "ensemble of opportunity"; Mote et al., 2011) to which adaptation may be needed. Specifically, climate scenarios that approximate smaller, median, and larger range of potential changes in precipitation intensity were selected for this analysis, using sets of two scenarios (one from RCP 4.5 and one from RCP 8.5) that appear to be near the 10[th], 50[th], and 90[th] percentiles of the distribution among GCMs of annual extreme precipitation in the Grand Rapids area. Ranking was evaluated based on both the largest daily extreme and average daily extreme across the 30-year time window. The analysis uses the 10[th] and 90[th] percentiles, rather than the most extreme models, as it is well recognized that some individual GCMs may provide unrealistic results for a given area. It is therefore standard practice to ignore the most extreme outliers and use a model at approximately the 90[th] percentile of the complete set of models as a reasonable upper bound. Use of such an upper percentile is generally considered appropriate for engineering/hydrologic design planning purposes; however, the full range may also be of interest.

For future time periods the analysis examined the mid-21[st] century and the end of the century. CMIP5 model projections run through 2100; however, they also incorporate decadal cyclical drivers of climate variability, so change evaluations can't be based on a single year. Therefore, 30-year time slices centered at 2050 and 2085 are used to represent mid- and late-century conditions.

An efficient method was needed for initial screening across 30 GCMs and 2 RCPs. As the primary interest was in maximum precipitation rates, not total monthly or annual precipitation volume, the initial screening was based on non-downscaled GCM results for maximum 1-day precipitation (by year and geographic location) extracted by the Expert Team on Climate Change Detection and Indices (Sillmann et al., 2013a; 2013b) and served as CLIMDEX by Environment Canada (http://www.cccma.ec.gc.ca/data/climdex/climdex.shtml).

This screening intentionally did not use measures of GCM skill or quality of performance in replicating past climate for the Midwest as primary criterion for model selection. In addition to the fact that performance on past climate is not necessarily a good predictor of future performance, Mote et al. (2011) have demonstrated that attempting to select GCMs based on performance does not appear to reduce uncertainty in projections. Some additional criteria were, however, used for selecting among the set of models near each of the percentile targets. Specifically, selection favored, where available, GCMs that use a finer spatial resolution or are among those with a better apparent fit in predicting summer precipitation (Toreti et al., 2013), both of which are likely to be associated with better prediction of summer convective storms. Some preference was also granted to GCMs demonstrated to have low global Bias (0.9 < B < 1.1) and a high volumetric hit index (VHI > 0.7) for mean monthly precipitation (Mehran et al., 2014). The selected GCMs are shown in Table 1.

**Table 1. GCMs Selected for Detailed Analysis at Grand Rapids, MI**

| Percentile | RCP 8.5 Models | RCP 4.5 Models |
|---|---|---|
| 90% | RCP8.5 MIROC5 (a) | RCP4.5 IPSL-CM5B-LR |
| *50%* | *RCP8.5 CNRM-CM5 (a, b)* | *RCP4.5 CNRM-CM5 (a, b)* |
| 10% | RCP8.5 bcc-csm1-1 (b) | RCP4.5 bcc-csm1-1 (b) |

Notes: a. High-resolution (Toreti et al., 2013); b. Low bias, high volumetric hit index (Mehran et al., 2014)

It should be noted that rankings on final analysis of the magnitude of storms of a given intensity and duration from downscaled GCM output may not follow the same order as shown in the CLIMDEX screening of non-downscaled GCMs. This occurs because the GCM output is subject to local bias correction during the downscaling process. As a result of this, a GCM that has lower peak rainfall intensity in the raw output could actually yield higher peak intensity in downscaled output if the bias correction factor is larger than that for other competing models.

Nonetheless, sampling across a range of GCM behavior is useful to help ensure an inclusive data set.  This suggests that final future IDF curves for application should be selected from the largest responses observed in the six model sample set described above, regardless of the ranking of the un-downscaled GCM, to provide reasonably protective design standards.

### *Spatially Downscaled Climate Data*

GCMs generate output at a large spatial scale (typically about 1°x1° or coarser) that does not take into account details of local geography and topography.  To be useful at the local scale it is necessary to undertake spatial downscaling.  Downscaling can be done either through the use of a small-scale regional climate model (RCM) or through statistical methods.  RCMs are computationally expensive to run, so only a limited number of GCMs have been downscaled in this way.  In contrast, there are many different varieties of statistically downscaled products now available.  Most of these work with the general design of using spatial statistical corrections of GCM monthly output to local spatial scales with bias correction based on analysis of GCM ability to replicate historical climatology, followed by temporal downscaling to a daily time step. A recent recommendation from the U.S. Fish and Wildlife Service (Patte, 2014) is to use the Multivariate Adaptive Constructed Analogs (MACA) statistically downscaled data (to a 4 km x 4 km scale) created by the University of Idaho.  The MACA method (Abatzoglou and Brown, 2012) has two advantages that make it slightly preferable to other downscaling methods: (1) it provides simultaneous downscaling of precipitation, temperature, humidity, wind, and radiation (rather than just precipitation and temperature), helping to ensure physical consistency in the outputs, and (2) the method uses a historical library of observations to construct the downscaling in the constructed analogs approach such that future climate projections are distributed from the monthly to the daily scale in comparison to months that exhibit similar characteristics in the historical record.  The approach therefore retrieved downscaled GCM output for the location of Grand Rapids International Airport (the observing station from which Atlas 14 results for the area are calculated) from the MACA website (http://maca.northwestknowledge.net/).  From these results the AMP series were extracted for calculation of updated IDF curves.

### *Results for Grand Rapids, MI*

The methods described above were applied to a test case for the City of Grand Rapids, MI to estimate the potential future range of IDF curves for the time periods around 2050 and 2085. IDF curves were calculated for 2, 5, 10, 25, 50, 100, 200, 500, and 1000-year events at variety of durations ranging up to 24 hours.  Results for the 25-year recurrence event for the 2085 time horizon are shown for example in Figure 1.  The uncertainty in future climate predictions is obvious as the results range from a 25 percent increase in the 25-year event (BCC-CSM-1-1, RCP 4.5) to an almost doubling of the 24-hour event (MIROC5, RCP 8.5).

As a check on the calculations, Kao and Ganguly (2011) suggest that change in precipitation extremes should reflect change in the potential saturated water content of the atmosphere in response to temperature change.  Based on the ideal gas law and the Clausius-Clapeyron relationship, saturated water content and precipitation extremes from time 1 to time 2 should scale as:

$$\frac{T_1 + 273}{T_2 + 273} \; x \; exp \left[ \frac{17.27 \, T_2}{237.3 + T_2} - \frac{17.27 \, T_1}{237.3 + T_1} \right] \quad \textbf{[15]}$$

where $T_i$ is air temperature in Celsius at time $i$.

The USGS National Climate Change Data Viewer (Alder and Hostetler, 2013; http://www.usgs.gov/climate_landuse/clu_rd/apps/nccv_viewer.asp) provides summaries of monthly GCM output from a full suite of CMIP5 GCMs focused to the Grand Rapids area (Kent Co., MI) based on 25 year time slices.  For the 2075 – 2099 time period, the predictions of maximum monthly average temperature under RCP 8.5 range from 31.4 to 39.8 °C, with a model mean of 34.8 °C, compared to the historical (2050 – 2005) maximum monthly average of 28.2 °C.  Monthly temperature is probably the most relevant measure for extreme event forecasting as precipitation rarely occurs on the hottest individual day in a month.  This suggest that potential precipitation extremes should increase by a factor ranging from 1.18 to 1.83, with a model mean of 1.42.  The statistical calculations described in this paper for the time slice centered at 2085 project an increase in the 24-hour precipitation with 25-year recurrence for RCP 8.5 with a median of 1.47 and a range of 1.42 to 1.62 times the recent historical record, in good agreement with the theoretical increase.

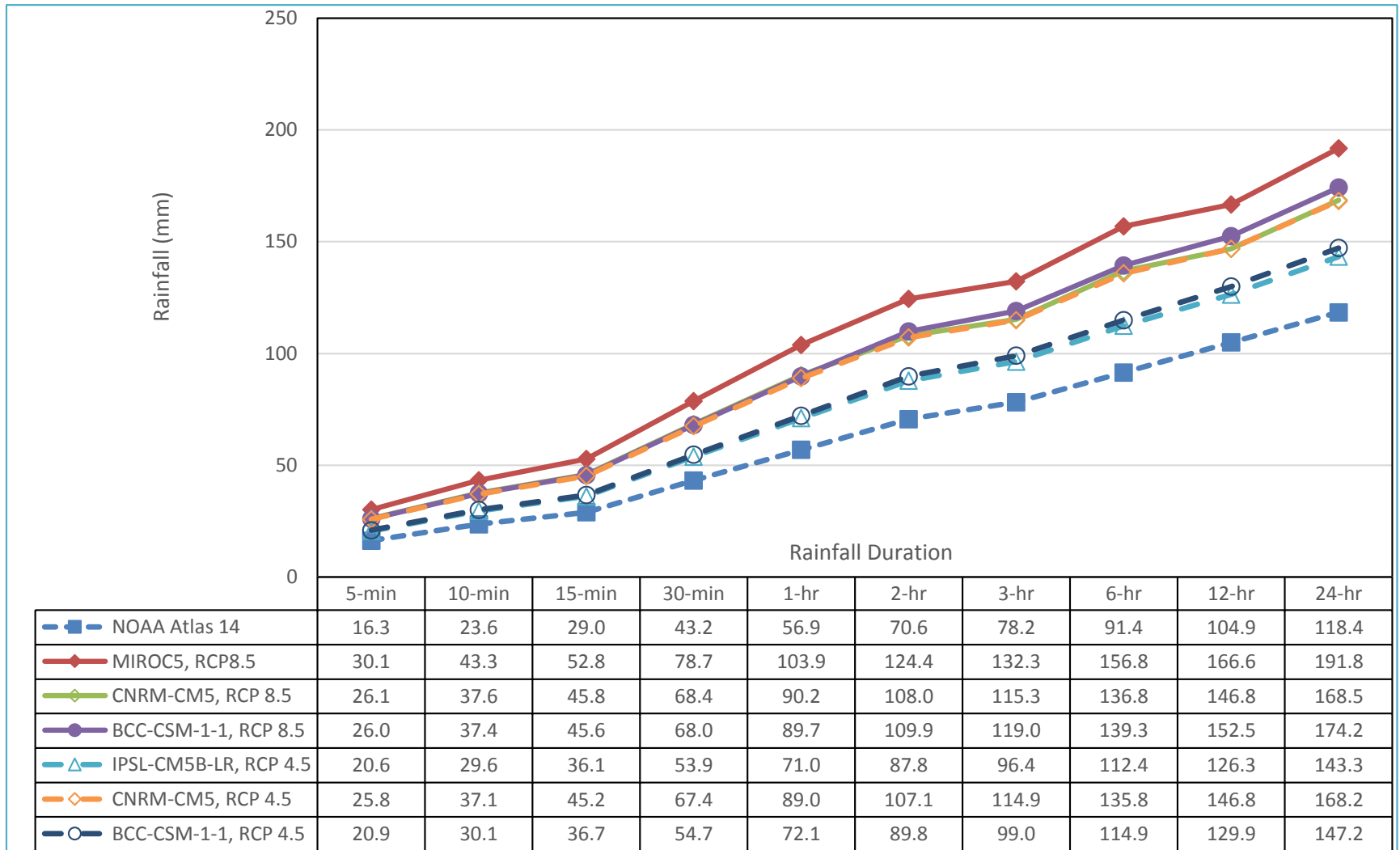

| | 5-min | 10-min | 15-min | 30-min | 1-hr | 2-hr | 3-hr | 6-hr | 12-hr | 24-hr |
|---|---|---|---|---|---|---|---|---|---|---|
| NOAA Atlas 14 | 16.3 | 23.6 | 29.0 | 43.2 | 56.9 | 70.6 | 78.2 | 91.4 | 104.9 | 118.4 |
| MIROC5, RCP8.5 | 30.1 | 43.3 | 52.8 | 78.7 | 103.9 | 124.4 | 132.3 | 156.8 | 166.6 | 191.8 |
| CNRM-CM5, RCP 8.5 | 26.1 | 37.6 | 45.8 | 68.4 | 90.2 | 108.0 | 115.3 | 136.8 | 146.8 | 168.5 |
| BCC-CSM-1-1, RCP 8.5 | 26.0 | 37.4 | 45.6 | 68.0 | 89.7 | 109.9 | 119.0 | 139.3 | 152.5 | 174.2 |
| IPSL-CM5B-LR, RCP 4.5 | 20.6 | 29.6 | 36.1 | 53.9 | 71.0 | 87.8 | 96.4 | 112.4 | 126.3 | 143.3 |
| CNRM-CM5, RCP 4.5 | 25.8 | 37.1 | 45.2 | 67.4 | 89.0 | 107.1 | 114.9 | 135.8 | 146.8 | 168.2 |
| BCC-CSM-1-1, RCP 4.5 | 20.9 | 30.1 | 36.7 | 54.7 | 72.1 | 89.8 | 99.0 | 114.9 | 129.9 | 147.2 |

2  **Figure 1.  Updated 2085 IDF Curves for 25-year Recurrence Precipitation at Grand Rapids, MI**

## 3  Discussion

Different GCMs predict a range of future precipitation regimes for Grand Rapids.  Depending on the model, total rainfall may increase or decrease.  It is clear, however, that warming air temperatures provide additional energy that is likely to increase the intensity of storm events.

This study provides methods for efficient calculation of future IDF curves based on the transform from GCM simulations of historic to future climate annual maximum series.  The method is illustrated by evaluation of a range of projected climate scenarios for Grand Rapids, MI that span from the low to high end of future precipitation intensity as a way to identify the range of conditions to which adaptation may be needed.  For 2050 conditions, one scenario (BCC-CSM-1-1, RCP 4.5) projects little change, and in some cases a decrease in rainfall intensity; however, the remaining scenarios show increases – up to 24% for the 24-hr 2-year event and up to 80% for the 24-hr 100-year event.  The models that predict the largest increases in intensity for a given duration may well be over-estimates for future conditions, but this cannot be ascertained in advance.  It is reasonable precautionary approach to use the upper bounds of the estimated IDF curves for 2050 as protective design standards for future conditions.  Additional changes are projected for 2085.

The results calculated include recurrence intervals of up to 1000 years for consistency with NOAA Atlas 14.  Estimates of extremely low probability events such as this are always subject to high levels of uncertainty.  For future climate conditions this uncertainty is amplified by questions about the ability of the climate models to resolve such rare events.  The longer-recurrence results are important in a qualitative sense to show how the risk of extreme flooding could increase, but it may be preferable to base quantitative design and management recommendations to results from the 100-year or lesser recurrence interval.

26  The methods described above can be largely automated through Python code and applied to

27  locations throughout the U.S.  Other design criteria, such as the 90[th] percentile 24-hour

28  precipitation event, can be analyzed in a similar manner.  The primary difference is that the

29  distribution of the 90[th] percentile event can be described by a Peaks-over-Threshold (POT)

30  approach, which characterizes the frequency of events greater than a specified magnitude

31  (Serinaldi and Kilsby, 2014).  As the value of the threshold ($u$) increases, the distribution of the

32  POT (prob $Y := (X\text{-}u)|X > u$) converges to a generalized Pareto distribution (GPD; Pickands,

33  1985; Balkema and de Haan, 1974):

$$H(y) = 1 - \left(1 + \xi \frac{y}{\tilde{\sigma}}\right)^{-1/\xi}, [16]$$

35  in which $\{y: y > 0 \text{ and } 1 + \xi \frac{y}{\tilde{\sigma}} > 0\}$ and $\tilde{\sigma} = \sigma + \xi(u - \mu)$, $\mu$ is the location parameter, $\sigma > 0$ is

36  the scale parameter, and $\xi$ is the shape parameter.  An updating procedure for the GPD, similar to

37  that described above for the GEV distribution, can be readily applied to estimate the distribution

38  of future 90[th] percentile events.

**Acknowledgments**

40  Funding for this work was provided by the City of Grand Rapids, MI, for which we offer thanks.

**Notation List**

42  *The following symbols are used in this paper:*

43  AEP    Annual exceedance probability

44  AMP   Annual maximum precipitation series

45  ARI    Average recurrence interval

46  CDF    Cumulative distribution function

| | | |
|---|---|---|
| 47 | c | Subscript for current climate results |
| 48 | EQM | Equidistant quantile matching approach |
| 49 | $F( )$ | Cumulative distribution function |
| 50 | $F( )^{-1}$ | Inverse cumulative distribution function |
| 51 | $f( )$ | Distribution function |
| 52 | GCM | General circulation model (alternatively, global climate model) |
| 53 | GEV | Generalized extreme value distribution |
| 54 | GPD | Generalized Pareto distribution |
| 55 | $H( )$ | Cumulative generalized Pareto distribution |
| 56 | m | Subscript for model results |
| 57 | o | Subscript for observed data |
| 58 | p | Subscript for predicted climate results |
| 59 | QM | Quantile matching approach |
| 60 | $\mathbf{R}_{i,p}^{j}$ | Ratio of rainfall extremes between future and current conditions for GCM $j$, rainfall |
| 61 | | duration $i$, and AEP of desired recurrence interval $p$. |
| 62 | S | Recurrence interval for durations less than one hour |
| 63 | $X_{max}^{GCM}$ | GCM-predicted AMP series |
| 64 | $X_{max}^{STN}$ | Historical AMP series |
| 65 | x | Generic climate variable |

| 66 | $Y_{max}$ | Bias-corrected future climate series of annual precipitation maxima |
| 67 | $\boldsymbol{\theta}$ | Parameters of fitted GEV distribution (shape, location, scale) |
| 68 | $\mu$ | Location parameter of generalized Pareto distribution |
| 69 | $\xi$ | Shape parameter of generalized Pareto distribution |